%% file: main.tex
\PassOptionsToPackage{pdfpagelabels=false}{hyperref}
\documentclass{cidr-2020}
\usepackage{xspace}
\usepackage{amsmath}
\usepackage{amssymb}
\usepackage{bbm}
\usepackage{bm}
\usepackage{calc}
\usepackage{float}
\usepackage{enumitem}
\usepackage[usenames, dvipsnames]{xcolor}
\usepackage{ctable}
\usepackage{xcolor}
\usepackage{blkarray}
\usepackage[caption=false,lofdepth]{subfig}
\usepackage{enumitem}
\usepackage{hyperref}
\usepackage{listings}
\usepackage{courier}
\usepackage{xcolor}
\usepackage{algorithm}
\usepackage{algpseudocode}

{}

\lstdefinestyle{mySQL}{ %
    mathescape=true,
    language=SQL,
    basicstyle=\small\ttfamily,
    deletekeywords={MIN, YEAR},
    morekeywords={LIMIT, GROUP, OPEN, CLOSED, SEMI, COPY, BY, ORDER, DESC, POPULATION, METADATA, WORLD, SAMPLE, MECHANISM, PERCENT},
    showstringspaces=false
}
\let\origthelstnumber\thelstnumber
\makeatletter
\newcommand*\Suppressnumber{%
  \lst@AddToHook{OnNewLine}{%
    \let\thelstnumber\relax%
  }%
}

\newcommand*\Reactivatenumber{%
  \lst@AddToHook{OnNewLine}{%
   \let\thelstnumber\origthelstnumber%
  }%
}
\makeatother
\newcommand{\name}{\textsc{Mosaic}\xspace}

\newenvironment{packed_enum}{
\begin{enumerate}
  \setlength{\itemsep}{1pt}
  \setlength{\parskip}{0pt}
  \setlength{\parsep}{0pt}
}
{\end{enumerate}}
\newenvironment{packed_grep}{
\begin{description}[leftmargin=10pt]
   \setlength{\itemsep}{1pt}
   \setlength{\parskip}{0pt}
   \setlength{\parsep}{0pt}
}
{\end{description}}

\begin{document}
\title{Mosaic: A Sample-Based Database System for Open World Query Processing}

\numberofauthors{1}
\author{
\alignauthor
Laurel Orr, Samuel Ainsworth, Walter Cai,\\ Kevin Jamieson, Magda Balazinska, Dan Suciu\\
       \affaddr{Paul G Allen School of Computer Science, University of Washington}\\
       \email{\{ljorr1, skainswo, walter, jamieson, magda, suciu\}@cs.washington.edu}
}
\maketitle
\begin{sloppypar}
\begin{abstract}
Data scientists have relied on samples to analyze populations of interest for decades. Recently, with the increase in the number of public data repositories, sample data has become easier to access. It has not, however, become easier to analyze. This sample data is arbitrarily biased with an unknown sampling probability, meaning data scientists must manually debias the sample with custom techniques to avoid inaccurate results. In this vision paper, we propose \name, a database system that treats samples as first-class citizens and allows users to ask questions over populations represented by these samples. Answering queries over biased samples is non-trivial as there is no existing, standard technique to answer population queries when the sampling probability is unknown. In this paper, we show how our envisioned system solves this problem by having a unique sample-based data model with extensions to the SQL language. We propose how to perform population query answering using biased samples and give preliminary results for one of our novel query answering techniques.
\end{abstract}


\section{Introduction}
\label{sec:introduction}
\input{introduction.tex}

\section{Motivating Example}
\label{sec:motivating_example}
\input{motivating_example.tex}

\section{Data Model}
\label{sec:model}
\input{model.tex}

\section{Query Evaluation}
\label{sec:system}
\input{system.tex}

\section{Marginal-constrained SWGs}
\label{sec:swg}
\input{swg.tex}

\section{Related Work}
\label{sec:related_work}
\input{related_work.tex}

\section{Future Work}
\label{sec:future_steps}
\input{future_steps.tex}

\section{Conclusion}
\label{sec:conclusion}
\input{conclusion.tex}

\end{sloppypar}

\small
\bibliographystyle{abbrv}
\bibliography{references}
\end{document}

%% file: introduction.tex
At its core, data science's goal is to answer questions about populations of interest. As many population datasets are either unaccessible (e.g., private medical data) or do not exist (e.g., all people living today), data scientists must use samples to approximately answer questions over these populations. Historically, data scientists accessed samples mainly by designing their own sampling scheme and manually collecting the sample for processing.

However, in recent years, the number of public data repositories has increased, e.g., Data.World, and government data has become more available. These sources give data scientists access to hundreds of samples, but they do not necessarily indicate how the samples were collected. This is problematic because samples can be arbitrarily biased which, if not corrected for, can lead to inaccurate analytical results.

Typically, samples are debiased using knowledge of how the sample was collected (sampling mechanism), but as the mechanism is not always available to data scientists, they are forced to develop alternative techniques to ensure accurate sample analysis. These alternative debiasing techniques vary across disciplines and are often customized to fit a specific type of data~\cite{zagheni2015demographic,lovelace2015evaluating,farooq2013simulation}. For example, in~\cite{lovelace2015evaluating}, the authors use an iterative reweighting algorithm to reweight a sample of individuals using known population aggregates. In~\cite{farooq2013simulation}, the authors learn probabilities from a sample and population aggregates and then use sampling to generate a representative population for later analysis.

In all of these cases, the scientists manually implement solutions hard-coded for a specific dataset in order to answer queries about some population of data. They do not use a database management system (DBMS) for query processing. DBMSs are built to efficiently answer questions about data and should be an ideal system for these data scientists, but they are not being widely used by them.

One reason DBMSs have not been more broadly adopted by the data science community is the closed world assumption taken by traditional DBMSs. This assumption states that any tuple not in the database does not exist. With sample analytics, this assumption is not valid. Sample analytics require the open world assumption where tuples not in the database could still exist.
 
In this vision paper, we propose a novel DBMS that automatically assumes an open world and allows scientists to query populations that do not fully exist in the database. This is a new problem as this influx of publicly available, arbitrarily biased data is recent, and we are not aware of other prototype systems that tackle open world data debiasing. In recent prior work~\cite{orr2020themis}, we explored a first, specific approach to this problem. In this paper, we paint a broader vision based on our experience with our prototype.

The challenge in building an open world database is that no general purpose solution exists to debias sample data without knowing the sampling mechanism. We will need to invent generalized techniques for debiasing. Further, an open world DBMS will need to provide an interface for users to ask questions over the debiased data.

In this paper, we describe our envisioned system, \name, named because it is built from a collection of samples (pieces of the population). \name builds on our prior work in~\cite{orr2020themis} and extends it in three main ways:
\begin{packed_enum}
\item It treats samples as first-class citizens, allowing users to define and ingest sample data.
\item It allows users to define and query populations of interest through the use of population metadata (e.g., population aggregate data).
\item It gives users the ability to choose how ``open'' it can be when answering queries (i.e. how much can it assume tuples are missing from the data). 
\end{packed_enum}

To achieve each open world feature, we make the following technical contributions:
\begin{packed_enum}
\item A sample-oriented data model with new keywords to the SQL query language so users can add, query, and modify samples and populations (\autoref{sec:model}).
\item Two open world query processing techniques utilizing samples and population metadata (\autoref{sec:system}, \autoref{sec:swg}).
\end{packed_enum}

%% file: motivating_example.tex
As a motivating example, inspired by~\cite{zagheni2012you}, take a data scientist who is using Yahoo! emails to estimate the number of migrants in European countries. As there are no existing debiasing systems, to correct for selection bias from Internet usage varying by country, age, and gender, the data scientist manually models the bias and uses maximum likelihood to fit her model to reported, ground-truth migrant statistics from Eurostat. After learning a corrected weight of each tuple, she asks various queries on the reweighted email sample.

To show the ease and versatility if she instead used \name, we show some example \name queries below but explain the semantics of these queries in \autoref{sec:model}. We assume Eurostat has information on the number of migrants per country and the number of migrants per email provider, and we leave out attribute declarations for space.
\begin{lstlisting}[style=mySQL,numbers=left,xleftmargin=3em,escapeinside=||]
CREATE TEMPORARY TABLE Eurostat;|\Suppressnumber|
$\textcolor{red}{\textit{\texttt{...Ingest Eurostat reports to Eurostat table}}}$|\Reactivatenumber|
CREATE GLOBAL POPULATION EuropeMigrants;
CREATE METADATA EuropeMigrants_M1 AS
  (SELECT country, reported_count
   FROM Eurostat); 
CREATE METADATA EuropeMigrants_M2 AS
  (SELECT email, reported_count
   FROM Eurostat); 
CREATE SAMPLE YahooMigrants AS
  (SELECT * FROM EuropeMigrants
   WHERE email = Yahoo);|\Suppressnumber|
$\textcolor{red}{\textit{\texttt{...Ingest Yahoo sample to YahooMigrants}}}$
|\Reactivatenumber|
SELECT SEMI-OPEN country, email, COUNT(*)
  FROM EuropeMigrants
  GROUP BY country, email;|\Suppressnumber|
-- UK, Yahoo, 20000
-- FR, Yahoo, 9000
-- ...
|\Reactivatenumber|
SELECT OPEN country, email, COUNT(*)
  FROM EuropeMigrants
  GROUP BY country, email;|\Suppressnumber|
-- UK, Yahoo, 20000
-- UK, AOL, 20
-- ...
\end{lstlisting}

At a high level, she creates a population of all European migrants (line 3) and adds the Eurostat report as metadata (lines 4-9). She then creates a sample relation representing migrants with Yahoo! email (lines 10-12) and then directly queries the population of all European migrants. \name automatically debiases the Yahoo! migrant sample using the Eurostat metadata.

The first query (lines 15-17) shows the number of migrants per country and email provider if \name just performs sample reweighting. As the sample has no information on non-Yahoo! email users, the only email provider is Yahoo!. The second query (lines 22-24) shows that \name can also use the metadata to generate entirely new, missing tuples, such as migrants who use AOL email.

%% file: model.tex
\begin{figure}[t]
    \centering
    \includegraphics[width=\linewidth]{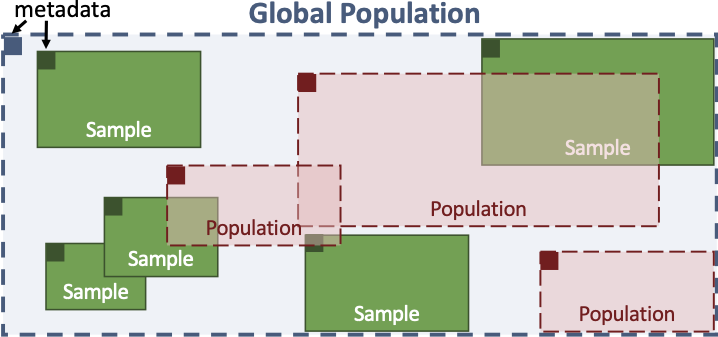}
    \caption{Data model of Mosaic.}
    \label{fig:model}
\end{figure}

We now describe the data model and main user interactions of \name. At a high level, \name defines two different specialized relations to help users analyze samples. In addition to these specialized relations, there is metadata associated with these relations to aid in analysis. Lastly, we define a new query construct to allow users to specify how much freedom \name has to reweight and create tuples when answering queries.

Before diving into the model, we need to define a key concept of sampling: the sampling mechanism. Given some reference global population being sampled, the sampling mechanism is the probability of an individual being included in the sample, denoted $\Pr_S(t)$.

A standard approach for sample analysis is to reweight the sample, meaning each tuple gets assigned a weight indicating how many tuples in the population that sample tuple represents. If the sampling mechanism is known, the stand approach is to reweight the tuples by $1/\Pr_S(t)$~\cite{beaumont2008new}.

\subsection{Relations}
There are three main types of relations in \name: population, sample, and auxiliary. We describe each relation and give our proposed syntax for declaring each of them.

\textbf{(1) (Global) Population Relation.} A population relation represents some set of tuples that could exists but are not fully known to \name. Populations are used for describing the set of tuples a scientist wants to query. When declaring a population, the user can use the \texttt{GLOBAL} keyword to enforce that this is the global population. The {\em global population (GP)} is the population that contains all other populations and samples and is used as a reference population when defining other populations and samples. Its use will become especially clear when doing sample reweighting in \autoref{sec:system}, and for now, we assume the user defines only one GP before defining other populations and samples.

A population is declared by
\begin{lstlisting}[style=mySQL]
CREATE [GLOBAL] POPULATION <pop> ($A_1$ type, $\ldots$)
  [AS (SELECT $A_1$, $\ldots$ FROM <gl_pop> WHERE <pred>)];
\end{lstlisting}
where there can be an arbitrary number of attributes. If the global keyword is not used, the population must be defined with a \texttt{SELECT} statement over a global population.

\textbf{(2) Sample Relation.} A sample relation represents a set of tuples that do exist in the GP and that \name has access to. A sample can have an associated sampling mechanism defined over the GP, but it is not required.

A sample is declared by
\begin{lstlisting}[style=mySQL]
CREATE SAMPLE <samp> ($A_1$ type, $A_2$ type, $\ldots$) AS
  (SELECT $A_1$, $A_2$, $\ldots$ FROM <gl_pop> [WHERE <pred>]
  [USING MECHANISM <$\texttt{mechanism}$> PERCENT <perc>]);
\end{lstlisting}
where the mechanism is some probability of inclusion defined over the GP, and the sample percent represents the size of the sample.

Some example mechanisms are \texttt{UNIFORM PERCENT 10} or \texttt{STRATIFIED ON $A_1$ PERCENT 20}. The former states that this sample is a 10 percent uniform sample, while the later states that this is a stratified sample on attribute $A_1$ where the sample contains 20 percent of the GP.

\textbf{(3) Auxiliary Relation.} Auxiliary relations are defined outside of any global population. They act the same as relations in traditional databases. In \name, they are primarily used for preprocessing, importing, and modifying data before ingesting it into \name's specialized relations. As these relations are created using the same constructs as standard SQL, we leave them out.

\subsection{Metadata}
\name uses specialized metadata for query processing. For sample relations, the metadata is tuple weights, initialized to be one for every tuple (the user can modify their initial values if desired). These weights are updated by \name's query engine (\autoref{sec:system:semi-open}) for query processing.

For population relations, the metadata is ground-truth information about the population. For example, the metadata could be covariances of various attribute combinations. While we envision a system the incorporates a variety of metadata sources, in this paper, we focus on using aggregate values for one or two attributes; i.e., 1- or 2-dimensional histograms. These histograms (marginals) are commonly released by corporations or governments to provide statisticians with population information while preserving anonymity (e.g., Data.Gov yearly reports). There is nothing preventing our system, however, from supporting higher-dimensional histograms. 

When \name answers queries over populations, it ensures these marginals are satisfied (\autoref{sec:system}).

The user specifies metadata via the command
\begin{lstlisting}[style=mySQL]
CREATE METADATA <relation> AS
 (SELECT $A_i$, [$A_j$], COUNT(*) FROM <aux_rel>
  GROUP BY $A_i$, [$A_j$]);
\end{lstlisting}
The user can update the initial sample weights via a similar command, but it is not required.

\subsection{Queries}
A core piece of \name's model is its query processing over populations. The challenge with answering queries over a population is that \name does not have access to all the population tuples.

To address this, \name adds a {\em level of visibility} to queries. The visibility of a query represents how \name can use the underlying samples. Motivated by the debiasing techniques discussed in \autoref{sec:introduction}, \name supports three main ways samples are accessed and used in analysis. (1) \name can answer queries using the samples directly without any attempt at debiasing. This is equivalent to standard data integration systems where the GP is treated as a global database, and the samples are local views over this database. (2) \name can reweight the samples to answer queries. Without the sampling mechanism, this is non-trivial. (3) \name can not only re-balance the samples but also infer and create missing tuples (e.g., inferring the number of UK migrants with AOL email). This allows \name to debias the sample and account for the fact that not all population tuples exist in the database.

When issuing a query, the user can decide which of the three visibility levels \name should use. The visibility level can be one of either \texttt{CLOSED}, \texttt{SEMI-OPEN}, or \texttt{OPEN} and appears after the \texttt{SELECT}. Closed means \name can only use samples (1), semi-open means it can reweight the samples (2), and open means it can reweight the samples and generate missing tuples (3).

The choice between \texttt{CLOSED} and \texttt{SEMI-OPEN} or \texttt{OPEN} represents if the user wants to make the open or closed world assumption. The choice between \texttt{SEMI-OPEN} and \texttt{OPEN} represents a trade-off between false-positive and false-negative tuples. In this setting, false-positives are tuples that \name returns as existing but are not (vice-versa for false-negatives). Specifically, if $n$ is the number of tuples existing in the population but are not present in the sample, then \texttt{SEMI-OPEN} will have $n$ false negative tuples but zero false positive. \texttt{OPEN} queries will potentially receive fewer false negatives but more false positives than \texttt{SEMI-OPEN}. The following table summarizes the trade-offs:
\begin{center}
\begin{small}
\begin{tabular}{|l|c|c|c|}
\hline
 & False Negative & False Positive & Assumption \\ \hline
\texttt{CLOSED} & $n$ & 0 & Closed \\ \hline
\texttt{SEMI-OPEN} & $n$ & 0 & Open\\ \hline
\texttt{OPEN} & $\leq n$ & $\geq 0$ & Open \\ \hline
\end{tabular}
\end{small}
\end{center}
The choice of query visibility level depends on the user's application and desired false positive and false negative rates.

We now describe how \name processes queries using metadata under the different levels of visibility.

%% file: system.tex
\begin{figure}[t]
    \centering
    \includegraphics[width=0.85\linewidth]{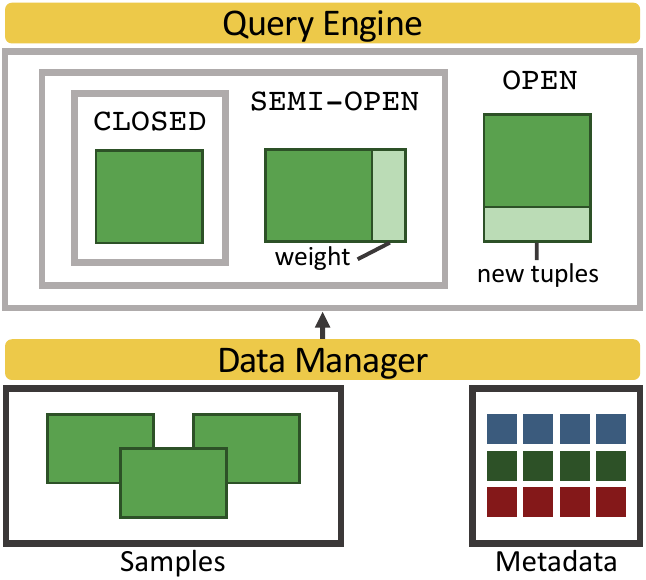}
    \caption{Proposed architecture of Mosaic.}
    \label{fig:architecture}
\end{figure}

\autoref{fig:architecture} shows the high level architecture of \name (we leave out auxiliary relations as they are standard SQL relations). The two types of stored data are the sample data and metadata. The user can interact with \name by either querying a population or updating the metadata.

To simplify the discussion, we make three assumptions:
\begin{packed_enum}
\item The population attributes are contained in the sample attributes (i.e., we do not need to join samples).
\item When a population query gets issued, the query engine receives a single, optimal sample to use (this can be relaxed by unioning samples over shared attributes).
\item There is metadata, i.e., marginals, about the global population or the query population.
\end{packed_enum}

As discussed in \autoref{sec:model}, \name answers queries using one of three different visibilities. As closed queries are the same as answering queries using views in a LAV data integration system, we refer the reader to~\cite{lenzerini2002data,halevy2006data} for query answering techniques and focus on semi-open and open queries. These two types of queries require novel system components, and in this section, we propose our solution that relies on marginals to answer arbitrary queries.

\subsection{Semi-Open Queries and Reweighting}
\label{sec:system:semi-open}
Recall that semi-open queries allow \name to perform sample reweighting. The two important subcases to consider are when the sampling mechanism is known versus unknown.

\begin{figure}[t]
    \centering
    \includegraphics[width=0.85\linewidth]{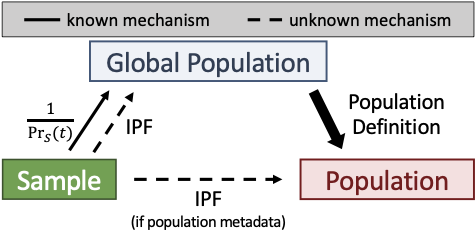}
    \caption{Semi-open world query evaluation.}
    \label{fig:semi-open_query}
\end{figure}

When the sampling mechanism is known, it is defined with respect to the global population. This is important as it guarantees that we can always approximately answer population queries. We use the known mechanism to reweight the sample by the inverse of its inclusion probability. Then we treat the population definition query as a view over this reweighted sample and run the user's query over this view. \autoref{fig:semi-open_query} shows this process.

\textbf{Research Question.} The main research question for semi-open queries is how to answer queries when the sampling mechanism is unknown using marginals. 

\textbf{Proposed Solution.} We addressed this problem partially in prior work when building \textsc{Themis}, an open world database system that automatically performs sample debiasing using population aggregates~\cite{orr2020themis}. \textsc{Themis} merges two techniques to answer simple count queries and group by queries. One approach performs sample reweighting using a technique called Iterative Proportional Fitting (IPF)~\cite{idel2016review, deming1940least}. The other approach builds a Bayesian network to represent the population probability distribution.


\name leverages the IPF technique presented in~\cite{orr2020themis} to answer arbitrary queries over samples. Specifically, we reweight the sample so that the given marginals are satisfied.

If there is metadata over the query population, \name uses IPF to directly reweight the sample to represent that population (bottom dashed line in \autoref{fig:semi-open_query}) and answers queries over the reweighted sample. If there is only metadata over the global population (left dashed line in \autoref{fig:semi-open_query}), \name reweights to represent the global population and then treats the query population as a view. Note that the accuracy will likely be lower when reweighting to fit global population (left dashed line) than reweighting to fit the query population directly (bottom dashed line) as biases that exist in the query population may not be captured when learning the global population.

\subsection{Open Queries and Population Generation}
\label{sec:system:open}
Open queries over a population allow \name to use a sample and marginals to generate or infer missing tuples (i.e., answer queries with fewer false negatives). We use the term infer because some aggregate queries can be answered without needing to materialize missing tuples.

At a high level, if we can learn the population probability distribution, we can either answer queries over this distribution directly or over data sampled from this distribution that is representative of the population. For example, if we model the probability distribution as a Bayesian network, we can answer \texttt{COUNT(*)} queries using direct inference over the network. However, for top-k or group by queries, we need to materialize a sample.

Generating data to fit some joint distribution requires a {\em generative model}~\cite{goodfellow2016nips} as it allows sampling from the distribution. Explicitly defined generative models define a parametric specification for the likelihood which is then solved for and sampled from. Implicitly defined generative models provide some process for generating samples without specifying the likelihood. For example, Bayesian networks and variational autoencoders~\cite{doersch2016tutorial} are explicitly defined models while generative adversarial networks~\cite{goodfellow2014generative} are implicitly defines ones.

The benefit of using an explicitly defined model is the potential for aggregate queries to be answered without having to materialize data; e.g., the Bayesian network \texttt{COUNT(*)} example above. The downsides, however, are that the explicit distribution will makes some, possibly incorrect, assumptions over the population. Bayesian networks, for example, assume the population satisfies the independent assumptions implied by the network. As we do not have access to the population data, we have no way of verifying if the model assumptions are accurate or not.

Therefore, in \name, we propose to take the alternative approach of using an implicitly defined generative model to learn the population as this requires no distributional assumptions. However, any generative model can be plugged in and used to answer open queries as long as it can be trained on sample data and marginals.

\textbf{Research Question.} The main challenge for learning a generative model is how to train one without access to the entire population. We only have a biased sample and 1-or 2-dimensional marginals, and we cannot just build a generative model over the sample as it is not representative of the population.

\textbf{Proposed Solution.} We propose to modify a Wasserstein generative adversarial network (WGAN) to use sample and marginal data rather than the entire population. Our motivation for modifying a WGAN is that if we project our marginals onto 1-dimension, the Wasserstein distance can be calculated exactly and efficiently (called the sliced Wasserstein distance~\cite{rabin2011wasserstein}). This eliminates the need for a discriminator network and turns our WGAN into a sliced Wasserstein generator (SWG)\footnote{Our SWG is similar to the one from \cite{deshpande2018generative}, except we do not use a discriminator to choose projections as we have multiple low dimensional marginals rather than the full dimensional population.}. We call our generator a marginal-constrained SWG (M-SWG) as it learns to generate population data that satisfies the marginals.

%% file: swg.tex
We now detail our proposed open query processing technique. We briefly review WGANs, describe our M-SWG, and give preliminary results. 
\subsection{WGAN Overview}
Generative adversarial networks (GANs) are comprised of two networks, a generator $G$ and discriminator $D$, pitted against each other in a game theoretic sense. The discriminator attempts to distinguish between data from the true distribution and samples from the generator. Meanwhile, the generator takes as inputs random variables, typically Gaussian or uniform, from some latent space and is tuned to output samples that fool the discriminator. Upon successfully training both networks, the generator will output samples that closely match the true data distribution.

\begin{figure}[t]
    \centering
    \includegraphics[width=\linewidth]{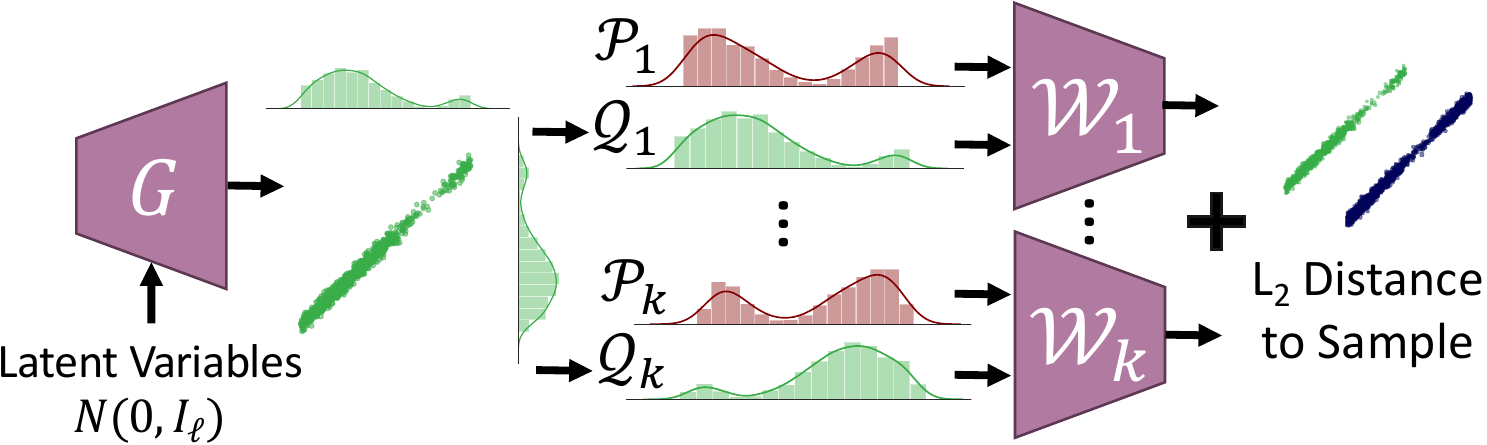}
    \caption{Marginal-constrained Sliced Wasserstein Generator.}
    \label{fig:gan}
\end{figure}

The Wasserstein or earth mover's distance, denoted $\mathcal{W}$, quantifies the optimal transport plan of moving ``mass'' between two distributions~\cite{wasserteindistance}. A Wasserstein GAN is a framework for training $G$ to generate data that minimizes the Wasserstein distance to the real distribution. Specifically, if $\mathcal{P}$ represents the real distribution and $\mathcal{Q}$ represents the generator distribution, WGANs solve
$$ \min_{G} \mathcal{W}(\mathcal{P}, \mathcal{Q}).$$
Although the discriminator does not appear in this formulation, \cite{arjovsky2017wasserstein} explains how the discriminator is used to calculate an upper bound on $\mathcal{W}$.

The problem with using traditional WGANs is they are trained with complete knowledge of $\mathcal{P}$. We need to modify WGANs so they can use the sample and marginals.

\subsection{Marginal-constrained SWGs}
We make two alterations to the standard WGAN. The first is to modify the optimization function to handle multiple marginals and a sample, and the second is to compute the discriminator exactly (turning our GAN into a generator network). Our proposed network structure is shown in \autoref{fig:gan}. Before discussing our solution, we need to define the assumptions of our generator model.
\begin{packed_grep}
\item[Manifold Hypothesis] The population and sample live in low dimensional $\mathbb{R}^{r}$ space but are embedded in higher-dimensional $\mathbb{R}^d$ space with $r \ll d$~\cite{narayanan2010sample}.
\item[Sample Coverage] For every point in our population, there is some point in our sample that is nearby.
\end{packed_grep}
The manifold hypothesis is a standard assumption made on high dimensional data and, in conjunction with the sample coverage assumption, means our sample informs us of the manifold our population data lives on. In particular, sample coverage requires our sample to give us the necessary structural information without needing to be distributionally correct. A stratified sample, for example, meets this criteria.

To more easily introduce our modifications, we first assume we only have 1-dimensional marginals, and then we extend our model to 2-dimensional ones. A subtle point is that if the population marginals do not cover all $d$ attributes (dimensions) of the sample, the model has no way of learning even the sample distribution of those attributes. Therefore, we add marginals from the sample into the set of population marginals for those uncovered attributes. We leave it as future work to determine the optimal set of 2-dimensional sample marginals to add in the 2-dimensional case.

\textbf{1-dimensional marginals.} Our first modification of the standard WGAN is to the error function. For notation, denote our sample as $S$, and assume we have an index set $\mathcal{I}_1$ of the 1-dimensional marginals. For each $i \in \mathcal{I}_1$, $\mathcal{P}_i$ represents the distribution of attribute $A_i$ of the population data (likewise for $\mathcal{Q}_i$). Our loss function becomes

$$\min_{G}\left[\sum_{i \in \mathcal{I}_1} \mathcal{W}(\mathcal{P}_i,\ \mathcal{Q}_i) + \lambda \mathbb{E}_{x \sim G}[\min_{y \in S}\|x - y\|_2]\right]$$

where $\lambda$ is a tuning parameter that trades off between fitting the population marginals and respecting the structure of the sample data. Note that $G$ takes as input random Gaussian variables of latent dimension $\ell$ where $\ell$ is treated as another tuning parameter.

The second modification we make is to compute the Wasserstein distance exactly~\cite{werman1985distance} instead of using the discriminator approach from~\cite{arjovsky2017wasserstein}. This modification is crucial to the practicality of our approach. Not only is computing $\mathcal{W}$ efficient for 1-dimensional data, but it makes the discriminator exact and avoids the need to train discriminator networks (one per marginal).

\textbf{2-dimensional marginals.} As we no longer have only 1-dimensional marginals, we cannot compute the exact Wasserstein distance efficiently as it requires solving a linear program for {\em every} step during training, which is prohibitively slow\cite{frogner2015learning,deshpande2018generative}. However, by using the sliced Wasserstein distance~\cite{rabin2011wasserstein,deshpande2018generative}, we can randomly project the marginals onto multiple one dimensional spaces and compute the Wasserstein distance exactly for each projection.

Specifically, in addition to $\mathcal{I}_1$, let $\mathcal{I}_2$ be the index set of the 2-dimensional marginals such that for each $\{i,j\} \in \mathcal{I}_2$, $\mathcal{P}_{i,j}$ represents the {\em joint} distribution of the attribute pair $A_i$ and $A_j$ (likewise for $\mathcal{Q}$). Further, assume we have a set of $p$ linear projections $\omega \in \Omega$ randomly generated and normalized to be on the unit sphere. Our loss function becomes
\begin{align}
\label{eq:training_error}
\min_{G} \Bigg[&\sum_{i \in \mathcal{I}_1}^{k} \mathcal{W}(\mathcal{P}_i,\ \mathcal{Q}_i) + \frac{1}{p}\sum_{\{i,j\} \in \mathcal{I}_2} \sum_{\omega \in \Omega} \mathcal{W}(\mathcal{P}_{i,j}\omega,\ \mathcal{Q}_{i,j}\omega) \\
&+ \lambda \mathbb{E}_{x \sim G}[\min_{y \in S}\|x - y\|_2]\Bigg] \nonumber
\end{align}
where $\mathcal{P}_{i,j}\omega$ is the 2-dimensional marginal over $A_i$ and $A_j$ projected by $\omega$. Note that while this discussion has focused on 2-dimensional marginals, the projection technique is applicable for arbitrarily higher-dimensional marginals.

\subsection{Preliminary Results}
We first evaluate just our M-SWG\footnote{Code available at \url{https://gitlab.cs.washington.edu/uwdb/project_mosaic_db}.} approach on a synthetic 2-dimensional population we can visualize. We then evaluate IPF (\name's \texttt{SEMI-OPEN} evaluation technique) and M-SWG on the flights data from~\cite{eichmann2018idebench} by running various aggregate queries. We compare against a uniformly reweighted sample which is the standard approximate query processing technique when there is no knowledge of how the sample was generated.

\begin{figure}[t]
    \centering
    \subfloat[{\small Biased sample and population.}]{
    \includegraphics[width=0.48\linewidth]{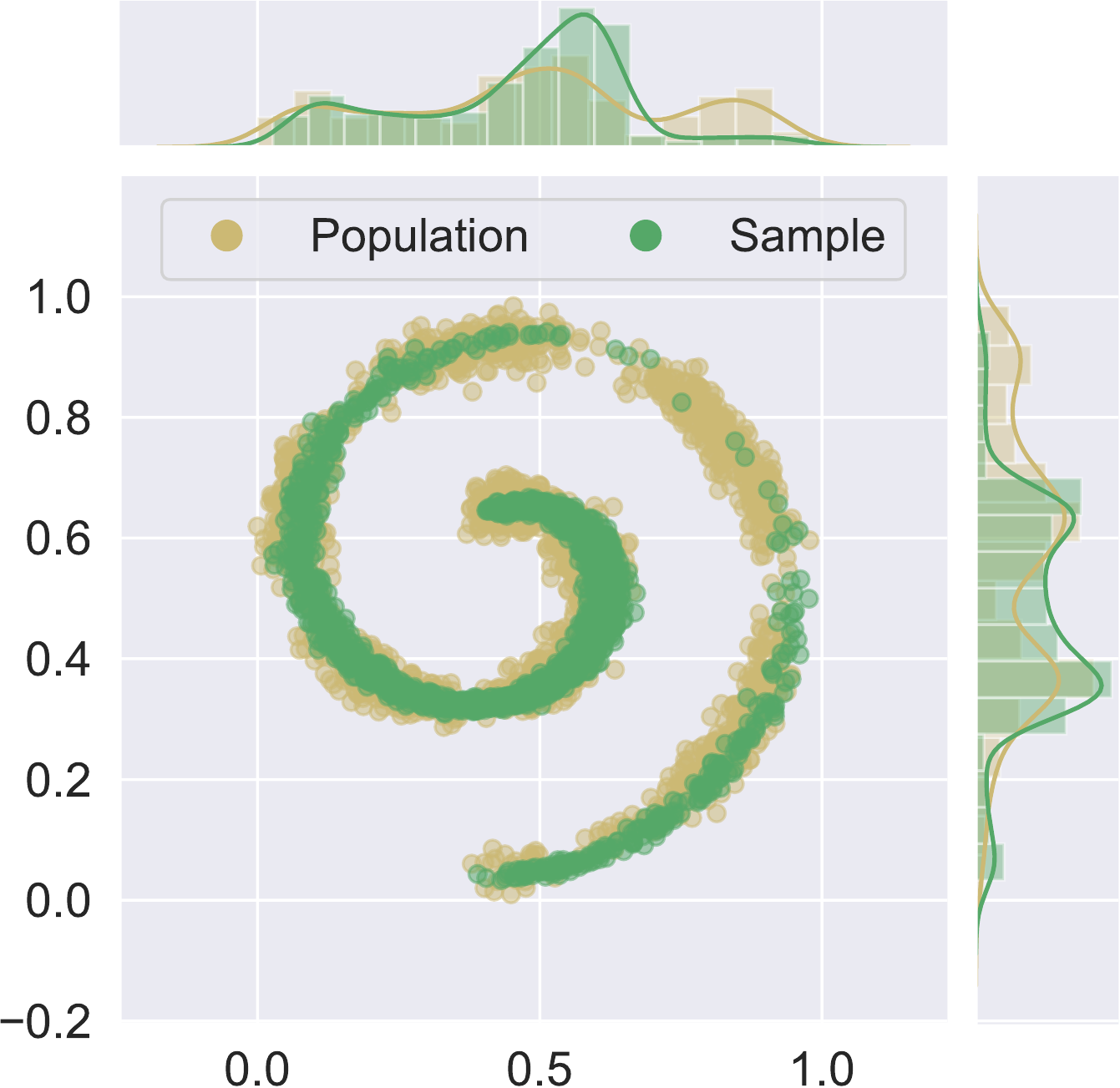}}
    \hfill
    \subfloat[{\small M-SWG sample and population.}]{
    \includegraphics[width=0.48\linewidth]{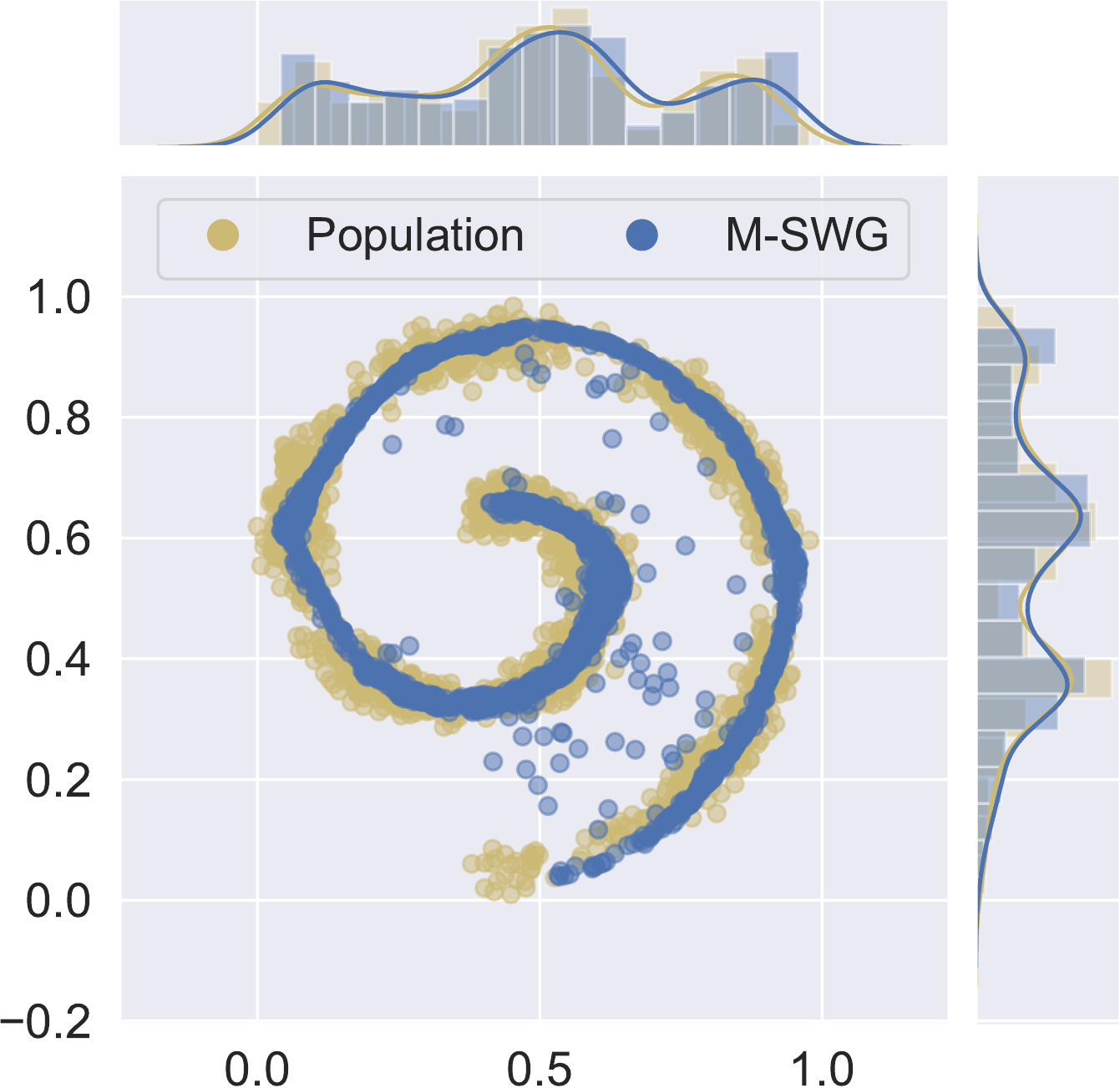}}
    \caption{Population data with original biased sample (a) and M-SWG generated sample (b).}
    \label{fig:joint_plots}
\end{figure}

\textbf{Synthetic Data.} We generate a 2-dimensional spiral population following the experiments from~\cite{cai2005learning} and generate a biased sample from this population with 10,000 rows (\autoref{fig:joint_plots} (a)). After training our M-SWG\footnote{We use 3 ReLU FC layers with 100 nodes each. We use $\lambda = 0.04$, $\ell = 2$, a batch size of $500$, and apply batch normalization after each layer. We use Pytorch's Adam optimizer with the default settings and an initial learning rate of $0.001$ that decreases by a factor of 10 if a plateau is reached during training.}, we use it to generate a sample also with 10,000 rows (\autoref{fig:joint_plots} (b)). Note that the generated data more closely matches the marginals while maintaining the spiral shape. Then, we uniformly reweight both the generated sample and the biased sample for query answering. We issue 100 random 2-dimensional range queries (i.e., counting the number of tuples lying in a bounding square box) for various box sizes. We repeat this for 10 different generated samples and report the average percent difference across the different samples.

\begin{figure}[t]
    \centering
    \includegraphics[width=\linewidth]{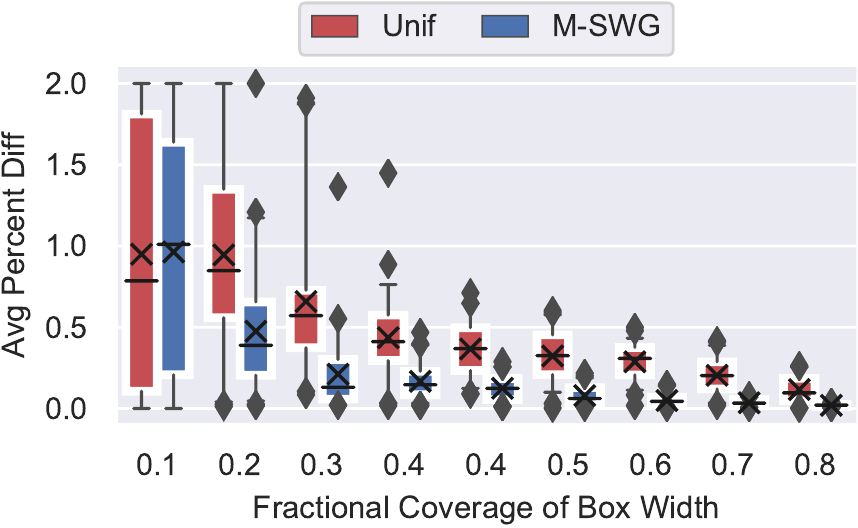}
    \caption{Average percent difference of uniform reweighting versus M-SWG for 2 dimensional queries.}
    \label{fig:results1}
\end{figure}

\autoref{fig:results1} show box plots (X is average) of the average query error for the 100 queries where the whiskers show the 3rd and 97th percentiles. We repeat this as we increase the size of our bounding box to cover a greater fraction of the data. For example, a width coverage of 0.8 means the range queries for 80 percent of the data on one dimension and 80 percent of the data on the other dimension. We see that we always outperform the uniformly reweighted sample except when the range is very narrow (asking for 1/100th of the possible values). For this highly selective query, both methods achieve high query error as both are more likely to have false negative tuples.

\textbf{Flights Data.} We use the flights data from IDEBench~\cite{eichmann2018idebench}, a real-world benchmark consisting of domestic flights in the United States (continuous attributes have been rounded to whole numbers), to explore the strengths and weaknesses of our approach. In particular, we study how IPF and M-SWG perform when we query numerical attributes only compared to \texttt{GROUP BY} queries with a single categorical group attribute. We use the attributes shown in \autoref{tab:flight_attr} and filter the data to just include flights in 2015 or 2016 (426,411 rows total) for faster processing. For M-SWG training, we one-hot encode the categorical variables and scale all attributes to be between 0 and 1. The attributes' encoded dimensionality (equivalent to the number of distinct values) is also shown in \autoref{tab:flight_attr}.

\begin{table}
\begin{footnotesize}
    \begin{center}
    \begin{tabular}{|c|c|c|}
    \hline
    \textbf{\texttt{Flights}} & \textbf{Abbrv} & \textbf{M-SWG Dim} \\ \hline
    \texttt{carrier} & \texttt{C} & 14 \\ \hline
    \texttt{taxi\_out} & \texttt{O} & 1 \\ \hline
    \texttt{taxi\_in} & \texttt{I} & 1 \\ \hline
    \texttt{elapsed\_time} & \texttt{E} & 1 \\ \hline
    \texttt{distance} & \texttt{D} & 1 \\ \hline
    \end{tabular}
    \end{center}
\end{footnotesize}
\caption{\texttt{Flights} attributes.}
\label{tab:flight_attr}
\end{table}

For our sample, we generate a biased 5 percent sample (21,320 rows) of flights with an elapsed flight time of more than 200 minutes with a 95 percent bias, meaning 95 percent of the tuples have a long flight time. For our metadata, we generate population marginals from the four attribute pairs of (\texttt{C}, \texttt{E}), (\texttt{O}, \texttt{E}), (\texttt{I}, \texttt{E}), (\texttt{D}, \texttt{E}). As the numerical attributes are already whole numbers, we do not need to build histograms, and the marginals are just projections of the population data.

\begin{table}[t]
\centering
  \begin{small}
  \begin{center}
  \begin{tabular}{|c|p{7.2cm}|}
    \hline
    \textbf{Id} & \textbf{Query} \\ \hline
    1 & \texttt{SELECT AVG(D) FROM F WHERE E > 200} \\ \hline
    2 & \texttt{SELECT AVG(I) FROM F WHERE E < 200} \\ \hline
    3 & \texttt{SELECT AVG(E) FROM F WHERE D > 1000} \\ \hline
    4 & \texttt{SELECT AVG(O) FROM F WHERE D < 1000} \\ \hline
    5 & \texttt{SELECT C, AVG(D) FROM F WHERE E > 200 AND C IN [`WN', `AA']} \\ \hline
    6 & \texttt{SELECT C, AVG(I) FROM F WHERE E < 200 AND C IN [`WN', `AA']} \\ \hline
    7 & \texttt{SELECT C, AVG(E) FROM F WHERE D > 1000 AND C IN [`WN', `AA']} \\ \hline
    8 & \texttt{SELECT C, AVG(O) FROM F WHERE D < 1000 AND C IN [`US', `F9']} \\ \hline
  \end{tabular}
  \end{center}
  \end{small}
\caption{The six SQL queries run in \autoref{fig:query_results}. We leave out the \texttt{GROUP BY} clause and replace \texttt{Flights} with \texttt{F} for space. `US' and `F9' stand for US Airways and Frontier (less popular carriers) while `AA' and `WN' stand for American and Southwest (more popular carriers).}
\label{tab:sql_queries}
\end{table}

Inspired by the queries from \cite{eichmann2018idebench}, we run the eight aggregate queries shown in \autoref{tab:sql_queries}. To run the aggregate queries over a weighted sample, we simple modify the aggregate to be over a weight attribute (e.g. \texttt{COUNT(*)} becomes \texttt{SUM(weight)}). To reduce the variance of the M-SWG approach, we generate $10$ samples with the same number of rows as the original sample and uniformly reweight the generated sample to match the size of the population. We return the groups appearing in all $10$ answers, averaging the aggregate value.

We add a softmax layer for the categorical variable and train our M-SWG similarly as above but with the latent dimension $\ell$ being the same as the input dimensionality. As our goal is an accurate representation of the data, we did not feel it necessary to experiment with reducing $\ell$. During training, we leave the softmax output continuous and only force the output to be binary for data generation.

We choose the model parameters by a small hyperparameter grid search\footnote{We search over the number of layers $=3,5,10$, number of hidden nodes $=50,200$, and $\lambda = 0.000001, 0.0000001$. When the number of hidden nodes is $200$ ($50$), we do not try $10$ ($3$) layers.}, running the models for three epochs (each epoch is one pass over the population marginals). We select the model receiving the lowest average query error from running 200 random queries over the continuous attributes with the same template as queries 1-4 where the attributes and predicates are randomly generated. We then rerun the chosen model with four different random initializations for 80 epochs and choose the one receiving the lowest error on the same 200 queries. Our final parameters are $\lambda = 1^{-7}$, $p = 1000$, $5$ layers with $50$ nodes each, and a batch size of $500$.

\begin{figure}[t]
    \centering
    \includegraphics[width=\linewidth]{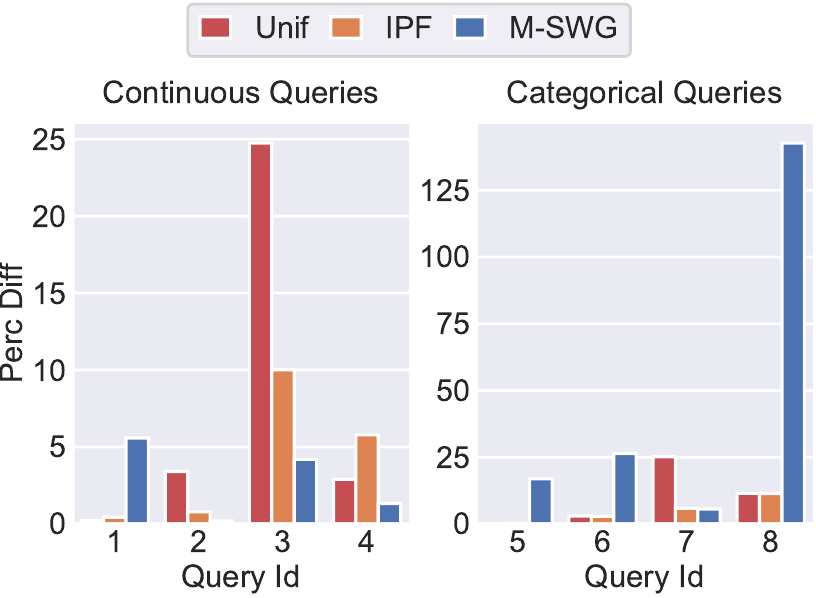}
    \caption{Average percent difference of uniform reweighting (default AQP) versus IPF versus M-SWG for queries 1-4 (left) and queries 5-8 (right). The y axis is on a different scale between the left and right plots.}
    \label{fig:query_results}
\end{figure}


\autoref{fig:query_results} (left) shows the average percent difference of the first four aggregate queries using uniform reweighting, IPF, and M-SWG. These are the queries over continuous attributes only. We see that all methods achieve an average error of less than 25 percent with M-SWG having the lowest max error of 7.5 percent. Averaged across all four queries, M-SWG gets the lowest query error compared to IPF and Unif.

Surprisingly, M-SWG receives the highest error on the query with a predicate matching how the sample is biased. IPF and Unif get almost no error for this query because the sample accurately represents this predicate. This is because M-SWG is slightly over-fitting to the population marginals and not generating as accurate a representation of the data satisfying this predicate as is in the sample.

IPF and Unif receive higher errors on query 3 because long distance flights are going to have longer flight times. As our sample is biased towards long flight times, the flight time is overestimated, even with a marginal over flight time and distance.

This trend of M-SWG receiving the lowest error on average for numerical queries is supported by the fact that on the 200 random queries used for parameter selection, when both the true answer and M-SWG answer are not-empty (161 queries total), {\em all} of our M-SWG models achieve a lower query error than Unif. IPF also achieves a lower error than Unif. The not-empty filter is due to the fact that M-SWGs struggles to capture light hitters (rare values) and will often not generate these rare values. This is because the error reduction in capturing the tail of a distribution (i.e. light hitter values) is often not significant enough for the generator to generate those values.


\autoref{fig:query_results} (right) shows the average percent difference of the last four aggregate queries using uniform reweighting, IPF, and M-SWG. Note that Unif and IPF get an close to zero error for query 5. For query 8, M-SWG receives a large error as it does not generate any flights with the carrier `US'. This is a result of the carriers attribute being categorical and having a skewed distribution in the data. Our M-SWG has to model an 18 dimensional space where the 14 one-hot encoded dimensions are sparse and not continuous (i.e., a one-hot encoded attribute is either a 0 or a 1). The M-SWG takes a continuous input, the latent variable $\ell$, and tries to generate data where 14 dimensions are binary and the distribution is skewed. This means there is not a significant reduction in error (\autoref{eq:training_error}) in generating the less popular carriers and potentially an increase in error from generating data falling between 0 and 1 for one of the binary dimensions, making it difficult to learn. We further discuss this problem in \autoref{sec:future_steps}.


These experiments have revealed that while our M-SWG approach is promising, there are still open questions around how to make it more accurate for heterogeneous, skewed data.

%% file: related_work.tex
Our vision relates to five main areas of research: data integration/exchange, approximate query processing, distribution learning, sample selection bias correction, and social science sample analysis. To the best of our knowledge, our prior work \cite{orr2020themis} was the first system to attempt automatic sample debiasing. \name is the first database to truly treat samples as first-class citizens in a declarative manner.

The database community has extensively researched data integration query processing (closed query processing in \name)~\cite{lenzerini2002data,halevy2006data,abiteboul2013complexity,kolaitis2018reflections}. While data integration systems do consider the difference between and the open and closed world assumption when answering queries over views, they do not reconstruct the global database. They are instead concerned with generating global schemas and giving equivalent or maximally contained answers. We go beyond data integration by providing the \texttt{SEMI-OPEN} and \texttt{OPEN} query constructs.

In a similar vein to \name's goal of supporting data scientists is the work on open data integration and exchange~\cite{miller2018open} where the focus is on helping scientists to discover and integrate publicly available data sources without apriori knowledge of what data is available. This work centers on finding relevant relations on the web~\cite{nargesian2018table} and schema mapping~\cite{popa2002translating}. We consider \name to work in tandem with these systems by providing scientists the tools to accurately analyze the data discovered from these open data sources.

While not concerned with data integration is the work of~\cite{ceylan2016open,friedman2019constrained} on open world probabilistic databases (PDB). Like \name, they also make an open world assumption but focus on theoretically understanding how different open world PDB semantics impact the precision and tractability of different classes of queries.

Approximate query processing (AQP)~\cite{chaudhuri2017approximate,mozafari2015handbook} explores how to achieve interactive query response times by allowing for approximate query results. Similar to \name, many AQP systems aim to accurately answer queries over samples for faster query processing~\cite{agarwal2013blinkdb,acharya1999aqua}. Further, some systems utilize past or known query answers to improve performance~\cite{peng2018aqp++,galakatos2017revisiting,park2017database}. The important distinction between \name and standard AQP is that AQP systems either assume they have access to the entire population or assume they have knowledge of the error from querying the sample. \name makes neither of these assumptions, making standard AQP techniques not applicable. However, if the user had knowledge of the error or assumed the error was normally distributed~\cite{galakatos2017revisiting}, these AQP systems could be used for \texttt{SEMI-OPEN} query processing in \name.

While most AQP research assumes a closed world, the work of~\cite{thirumuruganathan2019approximate,orr2019entropydb,kulessa2018model} allow for tuples to be generated that may not exist in the sample. Like \name, \cite{thirumuruganathan2019approximate} and \cite{kulessa2018model} also use generative models while \cite{orr2019entropydb} uses Maximum Entropy to learn a general probability distribution. Note that all systems assume access to the entire population, which does not hold in \name.

Most similar to \name's generator approach is~\cite{cai2005learning} where the authors learn the joint distribution of a mixture of Gaussians from lower dimensional projections of data. Sliced Wasserstein generators~\cite{deshpande2018generative} also learn a distribution from 1-dimensional projections of data. Further, \cite{liu2016coupled} use GANs to learn a joint distribution of images in two different domains by weight sharing across two GANs, one for each domain. However, all of these have the population so can take multiple projections. We only have marginals for specific attributes that may not cover all the attributes of the population.

Although not concerned with ad-hoc query answering, the research on sample selection bias correction in machine learning tackles the problem of training and test data being drawn from different distributions; i.e., the training data is biased and not representative of the true population distribution~\cite{cortes2008sample,huang2007correcting,bickel2007discriminative}. This work, however, assumes access to a sample from the true distribution (test set) and is mainly concerned with downstream tasks, such as classification or regression. \name solves the more general problem of debiasing samples for arbitrary queries using only population metadata and could be used to help machine learning scientists debias their samples.

Social science sample analysis, especially in demography research, is one of the most common use cases of biased sample analysis~\cite{muller2017generalized,deming1940least,zagheni2015demographic,fiorio2016migration,farooq2013simulation,lovelace2015evaluating,sun2015bayesian}. The samples are from social media or United States' surveys, e.g., the American Community Survey, and the aggregates for bias correction are census reports. While there are a few prepackaged libraries that perform specific algorithms, like the Python package IPFN, there is no general purpose technique or data management system for these scientists. Note that demography is not the only field that can benefit from \name, but it is one of our prime use cases and examples.

The database system most closely related to sample analytics for data science is MauveDB~\cite{deshpande2006mauvedb}. It allows model-based views over data to support users in building statistical models or abstractions over their data. While the overarching goal of supporting data science matches that of \name, the underlying mechanisms and data model is different.

%% file: future_steps.tex
We want to highlight that \name is a visionary system, and there remain numerous open challenges for \name to be fully realized. The following is a list of the more prominent challenges with ideas on solving them.

\textbf{Multiple Populations.} A main area of future work is to allow for multiple global populations. We not only need to keep track of which populations define the sampling mechanisms of which samples but also need to handle when samples exist in multiple populations and have a different sampling mechanism for each population. One possible solution is to treat all populations as disjoint and replicate any overlapping samples. However, we risk losing important debiasing information by this assumption. We leave it as future work to investigate ways of learning an optimal sample reweighting scheme from multiple mechanisms.

\textbf{Multiple Samples.} In \autoref{sec:system}, we made the assumption that the query engine receives a single sample for query answering. While simple, this is likely suboptimal as different samples represent different subsets of the population, and relying on one will result in more missed tuples. One solution is to union together all related samples and let IPF or the neural network reweight the tuples accordingly. This might present further problems if certain tuples become overly represented.

\textbf{Data Integration.} In \autoref{sec:system}, we also made the assumption that the population attributes are contained in the sample attributes. With real-world data, this is not the case. To relax this assumption, we will need to incorporate data integration query answering and schema mapping techniques to handle the disparate sample attributes. It is an open question whether standard data integration techniques hold or if they will need to be enhanced to handle the open world assumption of \name.

\textbf{Metadata.} Another aspect of future work is to add a greater variety of metadata. Our system already supports higher dimensional aggregate data, but it is an interesting direction to add metadata such as population constraints or known correlations.

\textbf{Open World Density Estimation.} Our SWG approach is simply one solution to open query processing. Based on \cite{thirumuruganathan2019approximate} and \cite{liu2016coupled}, \name's open query answering may be made more accurate by using more advanced sample generation approaches (e.g., ~\cite{azadi2018discriminator}) or by learning multiple SWGs on meaningful subsets of the data. Our SWG might also be made more accurate with a modified loss term to better ensure sample coverage. Further, it is an open question whether alternative density estimation techniques, like nonparamteric kernel density estimation~\cite{li2003nonparametric}, will be more accurate or efficient.

\textbf{Data Encoding.} The last research question is that of data representation. As mentioned in some of the (generative) model-based systems from \autoref{sec:related_work} and in \autoref{sec:swg}, it is unclear how best to represent heterogenous (continuous and categorical) data for query processing. The systems of \cite{ma2019dbest,orr2019entropydb,kulessa2018model} consider all possible attribute values when building models or answering queries. For small domains, this works well, but when the active domain gets large, these systems break down. \cite{thirumuruganathan2019approximate} takes a similar approach as \name by encoding categorical variables as binary numbers. While this reduces the dimensionality of one-hot encoding, it introduces various relationships between attribute values that may not exist. For example, suppose the three US states of CA, FL, and NY were encoded in two bits as $11$, $01$, and $10$, respectively. This implies that the sum of FL and NY is equivalent to CA, which is misleading. We think a promising approach is that of~\cite{marcus2019neo} where they used a word2vec model to learn {\em contextualized} encodings of values from the data.

%% file: conclusion.tex
\name is a DBMS that treats samples as first-class citizens and allows users to ask open world queries on biased samples of data and get approximate results. We described our proposed extensions to SQL to allow users to ask questions in a declarative manner and then discussed one possible way of doing open world query processing through the use of marginal data. In particular, we presented and gave preliminary results for our novel technique of using a M-SWG to generate population data.

While we focused on the use case of known population marginals (e.g. the census aggregate reports), the techniques presented are applicable for arbitrary or approximate marginals determined by the user. These marginals can represent any new distribution the user wants the sample to satisfy; e.g., one where classes are rescaled for downstream classification or where sensitive attributes are balanced. \name takes an important first step in building a system to better support the needs of data scientists by inherently supporting samples and the populations they represent.

\begin{small}
\noindent\textbf{Acknowledgments.} This work is supported by NSF AITF 1535565 and the Intel Science and Technology Center for Big Data.
\end{small}